# The sleeping bacterium: shedding light on the resuscitation mechanism

Eleonora Alfinito [1,*] and Matteo Beccaria [1,2,3]

1. Department of Mathematics and Physics 'Ennio De Giorgi', University of Salento, I-73100, Lecce, Italy;
2. National Institute for Nuclear Physics, INFN, Sezione di Lecce, via Arnesano, I-73100, Lecce, Italy;
3. National Biodiversity Future Center, Palermo 90133, Italy.
* Corresponding author: eleonora.alfinito@unisalento.it

## Abstract

The revival mechanism in dormant bacteria is a puzzling and open issue. We propose a model of information diffusion on a regular grid where agents represent bacteria and their mutual interactions implement quorum sensing. Agents may have different metabolic characteristics corresponding to multiple phenotypes. The intra/inter phenotype cooperation is analyzed under different metabolic and productivity conditions. We study the interactions between rapidly reproducing active bacteria and non-reproducing quiescent bacteria. We highlight the conditions under which the quiescent bacteria may revive. The occurrence of revival is generally related to a change in environmental conditions. Our results support this picture showing that revival can be mediated by the presence of different catalyst bacteria that produce the necessary resources .

## Introduction

Adaptation is one of the key principles to Darwinian theory of evolution. Among the mechanisms of adaptation, dormancy is one of the most fascinating and mysterious (Özgüldez et al. 2024): the organism enters a state of suspended life from which it can re-emerge when environmental conditions become more favorable. There are different definitions of dormancy (Özgüldez et al. 2024, McDonald et al. 2024), depending on the areas in which the phenomenon occurs, which have recently led to a multidimensional classification of the properties of this state/condition (McDonald et al. 2024). All definitions have in common the reversibility, i.e. the possibility of entering-exiting this state, and the reduced or absent ability to reproduce(Özgüldez et al. 2024, McDonald et al. 2024) . In recent years, dormancy has been often identified with the so-called VBNC (viable but not-culturable) condition (Oliver, 2005, Wagley et al. 2021), although as shown elsewhere, for example in (McDonald et al. 2024), this definition may seem somewhat reductive. On the other hand, VBNC has attracted much interest, especially in the case of pathogenic microorganisms, because it prevents their detection with conventional methods (Bari et al. 2013).

The formation and well-being of many types of bacterial colonies are ensured by a mechanism known as quorum sensing (QS), i.e. the coordination between different bacterial entities that regulates gene expression when a threshold number of bacteria is reached (Bassler 1999, Miller et al. 2001, Henkle et al. 2004, Waters et al. 2005, Higgins et al. 2007, Ng et al. 2009, Dandekar et al. 2012, Confort et al. 2013, Bruger et al. 2016, Bruger et al. 2021). The counting of bacteria occurs indirectly, through the production and reception of one or more types of molecules known as autoinducers (AI). On the other hand, QS does not only play a role in the growth of colonies or in the production of public goods, but also influences other phenomena such as, for example, the resuscitation of dormant cells. For example, in (Ayrapetyan et al. 2014) cell-free supernatants containing AI-2, a type of AIs, were used to resuscitate strains of *Vibrio vulnificus* in the VBNC state. Neither the addition of food nor cell-free supernatants without AIs were sufficient to resuscitate dormant cells. Thus, it was concluded that resuscitation is closely related to QS. Similarly, as observed in (Bari et al. 2013), the addition of autoinducers AI-2 and CAI-1 to dormant colonies of *Vibrio cholerae* significantly increased their potential for cultivation. Furthermore, the role of QS in the formation of subpopulation of non-growing, antibiotic-resistance of *Legionella pneumophila* has been highlighted in (Personnic et al. 2021).

In addition to biological models of dormancy in cells (Alnimr 2015, Russo et al. 2024, Zou et al. 2022, McDonald et al. 2024, Brueger et al. 2012, Pshennikova et al. 2022), there are several theoretical models of this phenomenon in the recent literature. Ref. (Carneiro et al. 2024) presents a review of most recent mathematical models used to describe the formation of biofilms with or without dormant cells, and a mathematical model of dormant cell formation in a biofilm is reported in (Chihara et al. 2015), while in (Nevermann et al. 2024) a revised version of the cellular automaton "game of life" is used to describe dormancy. Most of the models were developed to describe specific types of bacteria, due to their effects on human health: for example (Anwar et al. 2024) provides a review of mathematical models produced in about 50 years on all developmental stages (including dormancy) in *Plasmodium vivax,* (Chen et al. 2020) studies the kinetic equations for *Escherichia coli* in the VBNC state and several models explain dormancy in tumor cells (Page et al. 2005, Páez et al. 2012, Wilkie 2013, Mehdizadeh et al. 2021). On the other hand, to the best of the authors' knowledge, there is currently no theoretical model linking QS to dormancy.

In this paper, dormancy and the occurrence of revival emerge naturally within the framework of a model of QS previously developed by the authors (Alfinito et al. 2022, Alfinito et al. 2023, Alfinito et al. 2024). In particular, we will show how the switch between the dormant and viable (active) state can be described by the interplay of two metabolic parameters, namely assimilation rate and productivity. Finally, we will explore the conditions under which the presence of a small percentage of viable bacteria induces the resurgence of quiescent bacteria.

## Materials and methods
### Materials

We model the development of a bacterial colonies in a limited space ( e.g., a dish) using "agents". Each agent represents a bacterial "nucleus", i.e. a bacterial aggregate which has some specific metabolic features (assimilation rate $\sigma$, time of ageing $\tau$, minimal size of replication $Q_{min}$, productivity coefficient $\alpha$, etc, see Table A1). Each set of metabolic features defines a specific phenotype. A homogenous colony consists of agents with the same features ( single phenotype), while a mixed colony is made of agents with different features (multiple phenotypes). The rules introduced to describe the evolution of the colonies are inspired by the *Vibrio harveyi* (Alfinito et al. 2022, Alfinito et al. 2023, Alfinito et al. 2024) which has a well known QS circuitry (Henke et al. 2004, Bruger et al. 2016, Brueger et al. 2021). In our model, agents have two main aims: reproducing and producing public goods (PG), both for the benefit of the colony. A low production of public goods characterizes individuals who cheat. Reproducing is mainly related to the assimilation rate $\sigma$, while the production of PGs is primarily related to the productivity coefficient $\alpha$. These two actions are in tension because part of the nutrients which should be used for replication are directed toward PG production. By increasing $\alpha$ we obtain a higher production of PG but a reduced ability of the colony to reproduce, thus introducing a fitness cost. An initial seed of a single phenenotype with high productivity coefficient and low assimilation rate may continue to explore the landscape without replication, eventually dying due to senescence and mimicking the VBNC condition. Instead, in the opposite scenario of a phenotype with high assimilation rate and small productivity, agents will be rapidly reproducing and dying to the exhaustion of resources. This dynamic represents the behavior of viable bacteria in conditions of limited resource availability.

**Methods**

As mentioned in the introduction, many types of both Gram-positive and Gram-negative bacteria base many, if not all, of their activities on QS. In general, it reflects the ability of these organisms to coordinate themselves, producing effects that are also relevant on a macroscopic scale. In our model, QS is described as a long-range interaction among agents. The source of this interaction is the *sensing charge Q*, which is a measure of the agent's size (Alfinito et al. 2022, Alfinito et al. 2023, Alfinito et al. 2024). Each agent occupies a node in a regular grid and is equipped with a set of features that define its phenotype. In this study, different phenotypes correspond to different values of the assimilation rate, $\sigma$, and the productivity rate, $\alpha$. Communication between agents is mediated by the potential associated with long-range interaction, and resources are hierarchically distributed from agents with the highest potential to those with the lowest potential.

The analysis is performed using a stochastic procedure (detailed in the Appendix) that describes the temporal evolution of a seed of agents initially randomly distributed on the grid. At each iteration, each agent can connect with others: this occurs with a probability that increases with the total sensing charge of the network and decreases as α increases. After links are established, each agent receives an amount of sensing charge proportional to σ times the number of agents contacted.

Finally, at the end of each iteration, each agent may reproduce (i.e. divide into two equal parts with the offspring occupying an empty neighboring node) or jump to an empty neighboring node. In both cases, the selected node is the one with lowest potential among all neighboring nodes.

There are two possible outcomes of the initial evolution of the seed: a. the formation of a colony that fills the grid and consumes the total amount of resources; b. the progressive decrease in the initial number of agents, who (after wandering in the grid) were unable to reproduce and died of old age.

## Result

The main role of QS is to produce a successful cooperation among bacteria of the same strain. In the following we explore the conditions under which QS allows some initial seeds to build up the colony, i.e. to reproduce and overcome extinction due to senility. In the case of colony formation, the colony fitness, here given as the ratio between the final and initial number of agents $fitness = N_f/N_i$, becomes greater than 1. The procedure consists of analyzing the evolution of 60 different realizations (each produced by the same percentage of initial seeds, $f_0$). The seeds may belong to a single phenotype or two different phenotypes.

**Single phenotypes: when dormancy is convenient**

Dormancy is a state in which the bacterial strain is not able to reproduce, but can survive for extended periods, often migrating (Oliver, 2005, Wagley et al. 2021). This state is achieved in adverse conditions and when they improve, the bacterium may revive. In our model, dormant bacteria are described by agents that do not reach the minimum charge necessary to reproduce. Therefore, they migrate around the landscape, looking for positions in which the potential is the lowest thereby enhancing the chances to receive new charges. As a consequence, they tend to distribute quite uniformly in space. Finally, after a time τ they die due to aging. The dormancy condition is due to an excessively large ratio α/σ between the productivity and assimilation parameters. Within the set of chosen parameters (in particular, the network size) the assimilation rate has to be greater than 8 in order to allow the initial seeds to reproduce, also for the smallest value of α. As σ increases, the probability to reproduce grows, as described in Figure 1 where we report the percentage of realizations that remain in the dormancy condition at increasing values of α in the range ($10^{-4}$-50), for values of σ going from 9 to 100. We can observe that the curves become increasingly steeper as σ grows and that when the metabolic rate is larger than 12 the probability of dormancy at the smallest values of α, α= $10^{-4}$, is zero.

In conclusion: a single strain may enter the dormancy condition (100% of non-reproducing colonies) either when the assimilation rate σ decreases or when the productivity rate α increases. For each value of σ there exists a corresponding value of α that drives the entire colony to dormancy. Dormancy can also occur if the assimilation rate is not sufficiently high, even at the smallest values of α.

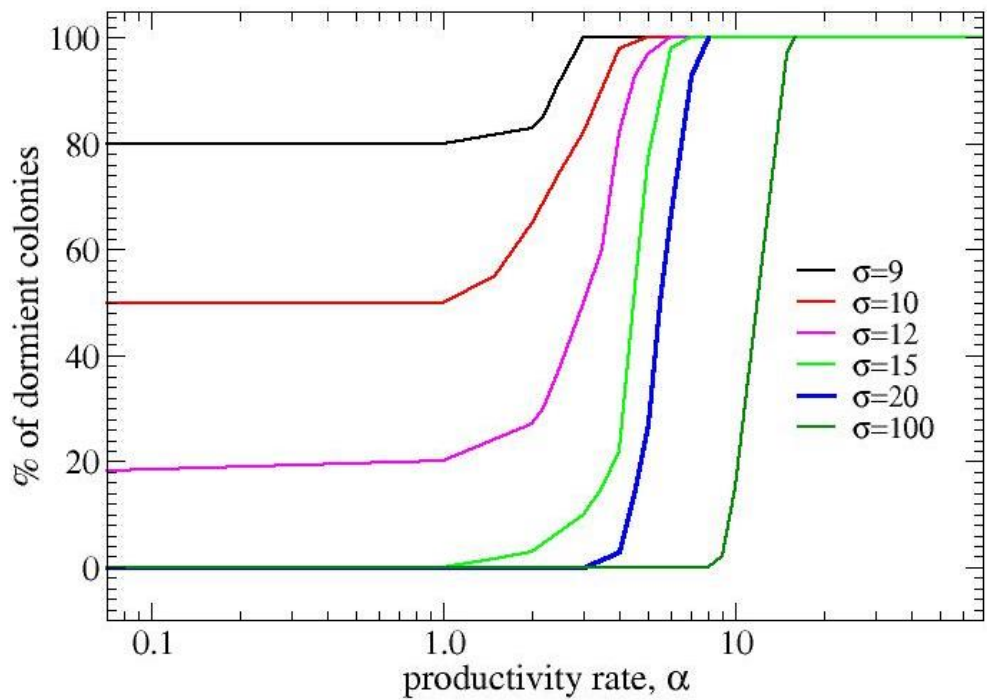

**Figure 1. Single-strain colonies.** Percentage of colonies in dormancy. Data correspond to strains with different values of the assimilation rate (σ) from 1 to 100, at growing vales of the productivity rate (α). Stochastic averaging is over 60 realizations. Grid size is 20x20 and the initial percentage of seeds is 10%.

**Two-strain colonies: revival of dormant bacteria**

A dormant phenotype (high productivity relative to the metabolic rate) can evolve in the presence of a rapidly growing phenotype (very low productivity relative to the metabolic rate). The first phenotype, referred to as dormant (D), is characterized by a long survival time (approximately twice the lifespan, $\tau$) during which agents migrate from initial positions in search of more favorable positions, eventually dying due to senility (Figure 2).

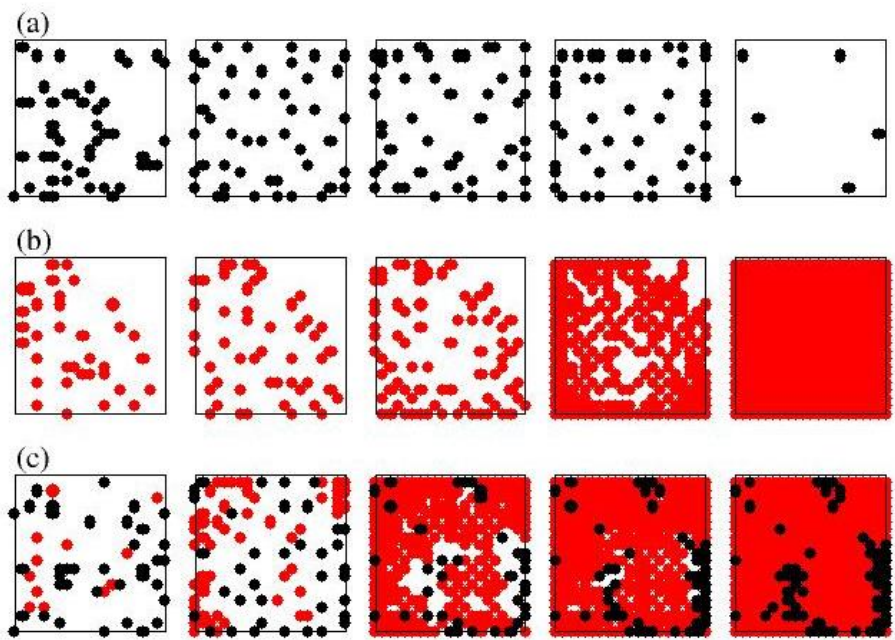

**Figure 2. Time evolution of pure and complex states**. Data concern: (row a) pure states low-metabolism agents (D); (row b) pure states high-metabolism agents (C); (row c) mixed states of both C and D agents. The initial percentages of agents are: 10% for the pure states and 7,5% for D and 2,5% for C in the mixed state. The assimilation rates ($\sigma$) are 2,20, for D and C, respectively, and the productivity rates ($\alpha$) are 2, $10^{-4}$ for D and C, respectively. Initial and final configurations (before extinction) are reported for each kind of state.

The second phenotype, hereafter referred to as catalyst (C), is instead characterized by a rapid growth leading to the complete filling of available space and the consumption of all available resources. When the two phenotypes are combined, we observe the formation of a mixed colony (CD) with a lifespan longer than that of the individual components; the phenotype previously called dormant reaches a fitness greater than 1, i.e. it becomes able to reproduce. The catalytic action of C is stronger when it is added in small quantities. However, if its initial concentration exceeds that of D, it quickly dominates the territory and does not allow the other to develop. Finally, when the productivity required for C is too high, it is no longer able to implement its catalytic action because it produces few charges and, finally, it ultimately fails to support D and the colony's lifespan approaches that of a single strain D. Figure 2 shows the evolution of a dormant ($\sigma_D = 2$, $\alpha_D = 2$) and a catalyst ($\sigma_C = 20$, $\alpha_C = 10^{-4}$) when left to evolve alone (Figures 2a, 2b) and mixed (Figure 2c). The initial concentration of the seeds is 10% for the single strains and 7,5% (D) vs 2,5% (C) for the mixed state. Note that, as anticipated in the previous section, type D bacteria tend to spread in space, occupying it almost uniformly.

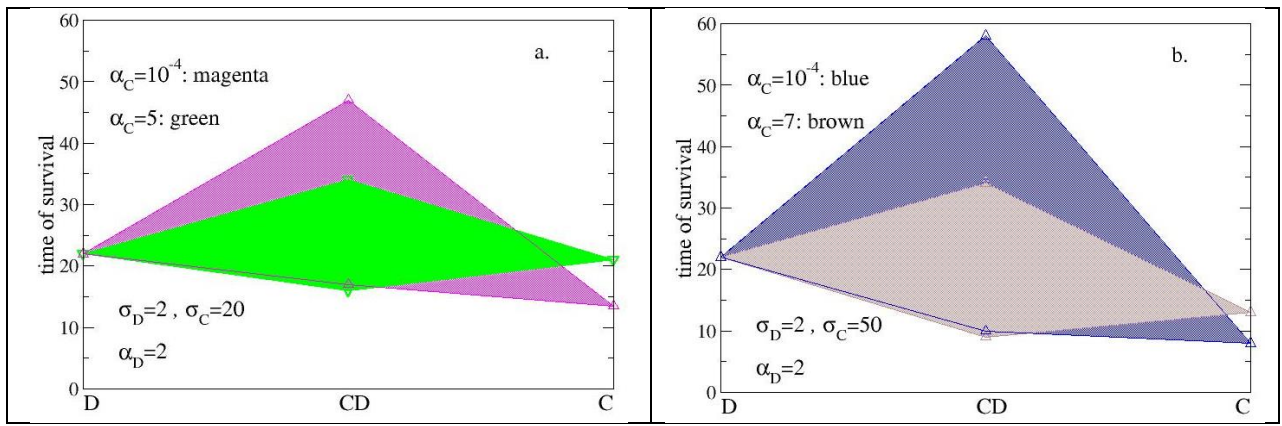

**Figure 3. Times of survival of mixed states**. Data concerns: single-strain colonies of low-metabolism cells ($\sigma_D=2$) (D); single-strain colonies of high-metabolism cells ($\sigma_C=20$) (C); mixed states of both (CD). The initial percentage of seeds, $f_0$, is 10% of the empty nodes. The survival time are obtained using 3 different initial percentages of the C strain: 2,5%, 5%, 7,5%. Going from the lowest to the highest initial percentages, the survival time of the mixed colony decreased. Data are reported in terms of *time-areas* are given for two different values of α: magenta ($\alpha_C = 10^{-4}$) and green ($\alpha_C = 5$). Time is calculated in iteration steps.

The survival time of a mixed colony depends, when all the metabolic parameters are chosen, on the initial concentration of the concurrent strains. In Figure 3 we report the colony survival time obtained using an initial concentration of the C strain from 2,5%, to 7,5%, with a total value of the initial concentration (C+D phenotypes) equal to 10%, $\sigma_D = 2$, $\alpha_D = 2$, $\sigma_C$, $\alpha_C$ variable. Data are presented in terms of *time-areas*, i.e. the area of a polygon whose vertices are: (from left to right) the survival times of the D and C phenotype alone, at the initial concentration of 10%; (from top to bottom) the survival times of the mixed colony obtained using the C-phenotype at the initial concentration of 2,5% and 7,5%. The larger the area, the more advantageous the mixed state is with respect the single state.

Both Figures 3a and 3b compare the role of different values of productivity for an assigned value of $\sigma_C$. Specifically, $\sigma_C$ is 20 in Figure 3a and 50 in Figure 3b: at increasing of the assimilation rate of the catalyst, the time-area increases. In particular, the largest time-area was obtained in both cases using a very small value of productivity ($\alpha_C = 10^{-4}$). In fact, the longest survival time increases, which correlates with an increasing of the fitness of the D-phenotype; the survival time of the C-phenotype alone becomes shorter (the strain consumes faster the available resources). The differences become smaller when the C-phenotype approaches the condition in which it reproduces with difficulty (more than 50% of the realizations are in the dormant state): this occurs using $\alpha_C = 5$ in Figure 3a and $\alpha_C = 7$ in Figure 3b.

To summarize: a small amount of fully viable phenotype (catalyst) can revive a dormant phenotype and this is a win-win condition because both gain something in terms of improved durability and fitness. Furthermore, revival depends on the metabolism and activity of the catalyst and is preferable for catalysts that have a very low propensity to produce public goods (small $\alpha$).

Finally, we complete our description of the colony formation by examining the evolution of the sensing charge $Q$. In particular, Figure 4 shows the distribution of charges of the D phenotype (top) and the C phenotype (bottom) for the evolution of a mixed colony. The top line is the complement of the bottom line (blue dots represent the cold positions, i.e. positions not occupied by the selected phenotype (C for the top line, D for the bottom line).

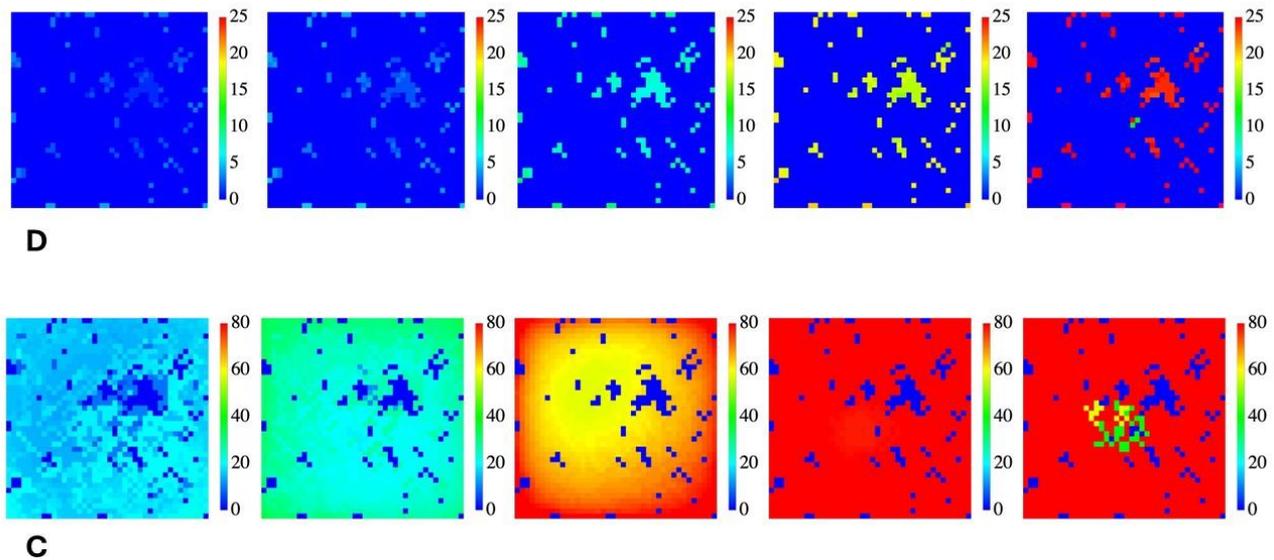

**Figure 4. Charge evolution of a mixed colony**. The color rainbow codifies the value of the sensing charge $Q$. On the top line the low-metabolism competitor (D), i.e. $\sigma_D=2$, on the bottom line the high-metabolism competitor (C), i.e. $\sigma_C=20$. The initial agents are given in equal concentration. The productivity rate, $\alpha$, is 2 for the D strain and $10^{-4}$ for the C strain. Pictures were taken at iterations 9,10,15,25, 32 for both phenotypes.

This figure gives us the opportunity to delve deeper into the meaning of the sensing charge $Q$ as a source of long-range interaction. As mentioned above, it represents the size of each bacterial aggregate and this is the core of the model: since we want to represent quorum sensing, the development of the colony is driven by the number of bacteria (hence by $Q$), the higher this value is, the more efficient the replication. On the other hand, it is important to remember that both Gram-positive and Gram-negative bacteria actually have a negative electrical charge on their cell surface (Wilhelm et al. 2021) and, the greater the bacterial aggregate, the greater the electric charge. Therefore, the sensing charge is a more general concept than just the electrical charge, although, like this, in this model, it plays the role of source of an action at a distance.

## Discussion and conclusions

The survival of weak organisms is linked to the ability to occupy an ecological niche in which they have no competitors. Weak living beings can coexist with stronger ones either by establishing a symbiosis with the strong ones or by prevailing over them. In both cases, it is not the individual that interacts with the stronger organisms but the community or the collective entity formed through the cooperation of the individuals. This phenomenon is well known in the case of bacteria that, if present in colonies (of many millions of individuals) can attack and overcome much larger organisms (for example using toxins (Miller et al. 2001, Higgins et al. 2007) or coexist with them (for instance, producing luminescence in cephalopods (Wei et al.1989)). These cooperative behaviour is rooted in the so-called quorum sensing, QS. The study of QS is therefore of significant interest both for limiting bacterial spread and, in general, for managing of benign bacteria.

One of the topics that is becoming increasingly relevant in the microbiology is that of potentially harmful dormant bacteria that could come back to life following climate change (for example those present in permafrost (McDonald et al. 2024)). A growing number of studies suggest a connection between dormancy and revival mechanisms in bacteria to quorum sensing. In the model proposed here, initially developed to measure bioluminescence in *V. harveyi*, the existance of a dormant state and the transition back to a viable state is controlled by two metabolic parameters. A strain can enter the dormant state in conditions of excessive demand for public goods (exogenous stress) or reduced use of available resources (endogenous stress). The presence of a second strain, the catalyst, can bring the dormant strain back to life, providing it with the nourishment it needs and implementing the QS mechanism that allows its use. In this case, the exit from dormancy is triggered by the presence of a reduced number of catalysts (too many of them, on the contrary, turn out to be too invasive to help). The improvement effect due to catalysts shows that they work best when they are less inclined to produce public goods. This behaviour corresponds to 'cheating' in the framework of sociomicrobiology (Parsek et al 2005, Passos da Silva et al. 2017). Thus, we conclude that even cheaters, as long as they are present in small quantities, can have a beneficial role for the purposes of the colony, or, in other terms, support the view that a healthy colony is made of individuals of different kinds (Bruger et al. 2016, Bruger et al. 2021).

As a final remark, dormancy is a hot topic not only in the field of microbiology, but especially in medicine, where it may concern the activation of cancer cells. Also in this case, dormancy seems to be an adaptive mechanism of survival for these cells, although it is unclear how and when they re-enter the cell cycle (Truskowski et al. 2023). In particular, at present there is no evidence that a coordination similar to QS exists also among cancer cells, although some form of cooperation, al least in metastatic condition, is presumable (Gkretsi et al. 2018, Hunter 2004). Further investigations in this field could be valuable for the prevention of recurrent tumors.

In conclusion, we have demonstrated how a simple model of hierarchical coordination (QS) in bacterial colony development accounts for the dormancy phenomenon. Ongoing investigations aim to further explore the role of other metabolic parameters included in the model, as well as their scaling with the size of the environment grid.

**Fundings:** This research did not receive any funding

**Data Availability :** All data generated or analyzed during this study are included in this article.
**Declarations**
**Conflict of interest :** The authors confirm that this article content has no conflict of interest.
**Author's contribution:** All the authors equally contributed to this reseach.

# Appendix

The procedure of colony formation is summarized in the following steps (Alfinito et al. 2022, Alfinito et al. 2023, Alfinito et al. 2024):

1. Start: on the assigned $N = L_x \times L_y$ grid, an initial amount, correspondent to the fraction $o_f$, of agents is randomly distributed. In the present investigation, $L_x = L_y = 20$, $f_o$ goes from 0.025 to 0.1. The agents may represent a single phenotype (all with the same features described in Table A1) or different phenotypes. Only one agent may occupy a node, the node is *active*

2. Colony formation: the following procedure continue until the total amount of energy $Max(E)$ is consumed or no more agents are present, due to too long time passed without replication, ageing time, ($\tau$)

2a. the energy and potential of each active node are calculated as follows:

$$V(l) = \sum_{j \neq l}^{N} \frac{Q(j)}{Dist(j,l)}, \quad \varepsilon(l) = Q(l) \sum_{j \neq l}^{N} \frac{Q(j)}{Dist(j,l)}, \quad (1)$$

where $Dist(j, l)$ is the Euclidean distance between the two nodes $j$ and $l$. $Q(l)$ is the *sensing-charge* of the *l*-agent

2b. the value of the total energy $E = \frac{1}{2} \sum_{\substack{i,j=1 \\ (i \neq j)}}^{N} \frac{Q(i)Q(j)}{Dist(i,j)}$ is updated.

2c. Each active node try to connect with other active nodes having a smaller value of potential. Connection happens with an assigned probability:

$$p(n,m) = \min(1, \exp(-\alpha^3 \Delta E_{n,m})), \quad with \quad \Delta E_{n,m} = \frac{\varepsilon(n) - \varepsilon(m)}{E} \quad (2)$$

Where $\alpha$ is the productivity index. In the case of multiple phenotypes, $\alpha$ represents the mean value of the specific $\alpha$ values.

Each new connection between the nodes $(n, m)$ produces the link $L(n, m)$ and the total number of links for the node $n$ is: $links(n) = \sum_m L(n, m)$

2d. Each node gains sensing charges proportionally to the value of the assimilation rate, $1 < \sigma < N$:

$$Q(n) \rightarrow Q(n) + floor\left(\frac{\sigma * links(n)}{N}\right), \quad (3)$$

2e. Each agent may colonize one of the nearest neighbouring nodes if its charge is larger than $Q_{min}$: this happens giving to the offspings half of its charge. Otherwise the agent migrate in the closest empty node with lowest potential.

This procedure is resumed in the flow-chart shown in figure A1. The code is available on request.

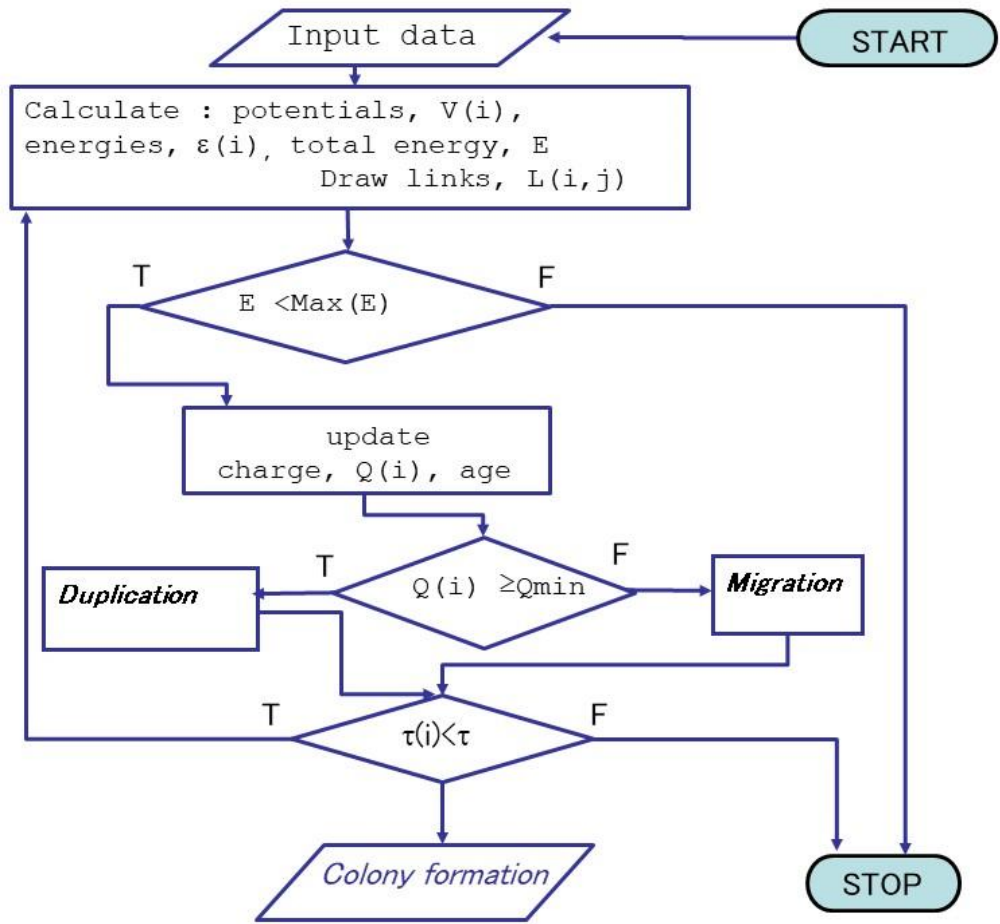

**Figure A1.** Flow chart of the algorithm describing colony formation

**Table A1.** Metabolic characteristic of the agents. Table resumes the symbol used in the text, their meaning and the values adopted in simulations.

| Symbol | Quantity | Range of values |
|---|---|---|
| α | Productivity coefficient | $[10^{-4}-7]$ |
| σ | Assimilation rate | [1-50] |
| $Q_{max}$ | Maximum value of the sensing charge | 80 |
| τ | Ageing time | 10 (a.u.) |
| Max(E) | Maximal fraction of energy to be used | 0.9 |
| $Q_{min}$ | Minimal replication size | 2 |